\documentclass[letter]{aa}  

\usepackage{graphicx}
\usepackage{txfonts}
\usepackage{natbib}
\usepackage{color}
\usepackage{fancyvrb}
\graphicspath{{./fig/}}

\title{Effects of granulation on the visibility of solar oscillations}

\author{J. Schou}

\date{\today}

\institute{
Max-Planck-Institut f\"ur Sonnensystemforschung, Justus-von-Liebig-Weg
3, 37077 G\"ottingen, Germany\\
\email{schou@mps.mpg.de}
}

\begin{document} 

 
\abstract
{
The interaction of solar oscillations with near surface convection is
poorly understood. These interactions are likely the cause of several problems
in helio- and astero-seismology, including the so-called surface effect and apparently
unphysical travel time shifts as a function of center to limb distance.
There is thus a clear need for further theoretical
understanding and observational tests.
}
{
The aim is to determine how the observed modes are affected by the convection.
}
{
I use HMI velocity and intensity images to
construct k-$\omega$ diagrams showing how the oscillation amplitude
and phase depend on the local granulation intensity.
}
{
There is a clear and significant dependence of the observed properties
of the oscillations on the local convection state.
}
{}


\maketitle
%

\section{Introduction}
The effect of convection on wave propagation is a complex issue.
With no clear separation of scales in time or space,
simple approaches do not work well.
This is particularly unfortunate as there are observed effects for the
Sun and other stars which are suspected of being due to such effects.
While attempts at more complete treatments
have been made
\citep[e.g.][]{2004A&A...418..305S,2013ApJ...773..101H,2015ApJ...806..246B},
the details are still far from understood.

One example is the so-called surface effect, which has been known for a long time \citep[e.g.][]{1988Natur.336..634C} and which manifests itself as
a difference between the theoretical and observed mode frequencies and
has properties suggesting a near surface origin. For recent discussions of this
see \cite{2014MNRAS.437..164P} and \cite{2014A&A...568A.123B}.

Another troubling effect is a time shift observed in
time-distance helioseismology with an apparently unphysical dependence
on the observable and a strong center-to-limb
dependence \citep[see, e.g.][]{2012ApJ...749L...5Z}.
An attempt to explain this was made by \cite{2012ApJ...760L...1B},
who noticed that eigenfunctions in hydrodynamic simulations exhibit
strong phase variations with depth and were able to roughly explain this, as well
as the center-to-limb effects, using a crude theoretical model incorporating some of
the properties of the convection.

Here I address the latter of these two issues by investigating
how the appearance of the oscillations depends on whether one
observes them in the middle of granules or in intergranular lanes.

\section{Method}
While the interaction of waves with convection is complex, it is probably
reasonable to try a simple model and start by considering a horizontally
uniform and initially vertically propagating wave in a Cartesian box.
The two largest contributors to the
distortion of the observed wave are likely the local thermodynamic state (i.e. sound
speed, as given by the temperature and thus intensity) and the vertical flow,
which are highly correlated.
Here I use the intensity, as it is easily measured and is less affected by
the modes than the Doppler velocity.
It is probably also reasonable to linearize the variations, that is to
assume that the observed velocity varies linearly with intensity.
Of course, the effect is not expected to be a real multiplicative factor, nor independent
of frequency, so is best described in the Fourier domain:
\begin{equation}
\widetilde V(\omega,I^\prime)=\widetilde V_0(\omega) (1 + \alpha(\omega) I^\prime)
\label{eq1}
\end{equation}
where $\widetilde {~}$ indicates Fourier transform,
$V$ is the line-of-sight velocity,
$V_0$ the unperturbed velocity,
$I^\prime$ is the intensity minus the mean (over time and space),
$\omega$ the frequency,
and $\alpha$ a complex factor to be determined.

To measure the effect I use data cubes from the Helioseismic and Magnetic Imager
\citep[HMI;][]{2012SoPh..275..229S} of Doppler velocity ($V$)
and computed continuum intensity ($I$), tracked by \cite{2014A&A...570A..90L}
at the rate of \cite{1984SoPh...94...13S} evaluated at the center of each cube.
These have 512 by 512 pixels of size 0.0005 radians by 0.0005 radians ($\approx$ 348 km by 348 km)
and 1920 points spaced by 45s in time, for a total cube size of $\approx$
180 Mm by 180 Mm by 24 hours.
For the present study cubes centered on latitudes $0^\circ$, $\pm 20^\circ$,
$\pm 40^\circ$ and $\pm 60^\circ$ crossing the central meridian near the center
of the time interval are used.
A few missing frames each day are filled by interpolating the adjacent
images linearly.
Also, for ease of interpretation, $I^\prime=I/{\bar I}-1$,
where $\bar I$ is the mean (over time and space) of $I$, is used.

Each frame of the cubes is subdivided into 4x4 pixel areas and two
quantities are calculated: $\bar V$, which is the average
of $V$, and $\bar S$ which is the slope from a linear fit of $V$ versus $I$.
Rather than oversampling, the resulting images have 
1/4 as many pixels in each direction.

In reality $\alpha$ might depend on the horizontal wavenumber of the wave,
so instead of simply averaging horizontally,
the resulting cubes are circularly apodized between 0.85 and 0.95 of their
half width with a raised cosine and Fourier transforms
$\widetilde V$ and $\widetilde S$
are calculated
from $\bar V$ and $\bar S$.
From these, $\alpha$ is obtained from the ratio of the cross spectra and power spectra as:
\begin{equation}
\label{eq2}
\alpha (k,\omega)= \frac{\langle\widetilde S (k_x,k_y,\omega,i) {\widetilde V}  (k_x,k_y,\omega,i)^* \rangle}{\langle\widetilde V  (k_x,k_y,\omega,i) {\widetilde V}  (k_x,k_y,\omega,i)^* \rangle},
\end{equation}
where $\langle\rangle$ denotes averaging over azimuth $\phi$ and time
(indexed by the cube number $i$)
\footnote{The averaging over $\phi$ and time is done by
first interpolating each of the cross- and power-spectra to a grid in $k$
and $\phi$, averaging over $\phi$ and then averaging over time.} ,
$k_x=k\cos\phi$, $k_y = k\sin\phi$,
$k=l/R_\odot$ is the (horizontal) wavenumber, $l$ the spherical harmonic degree,
$^*$~indicates complex conjugation,
and $R_\odot$ the solar radius.
For the results shown here the time averaging was over 30 one day cubes starting on 2010 June 1.

When determining $\bar V$ and $\bar S$ it is effectively assumed that the horizontal wavelength is long compared to 4 pixels, which may lead to some degradation of the results near the spatial Nyquist frequency of the $\bar V$ and $\bar S$ maps. However, far from the Nyquist frequency this should not be a problem.
Similarly, the use of $\bar V$ as an estimate of $V_0$
should be a good approximation as the term $\alpha(\omega) I^\prime$ in Eq. \ref{eq1} is small.

\begin{figure}[t!]
\begin{center}
\includegraphics[width=0.9\columnwidth]{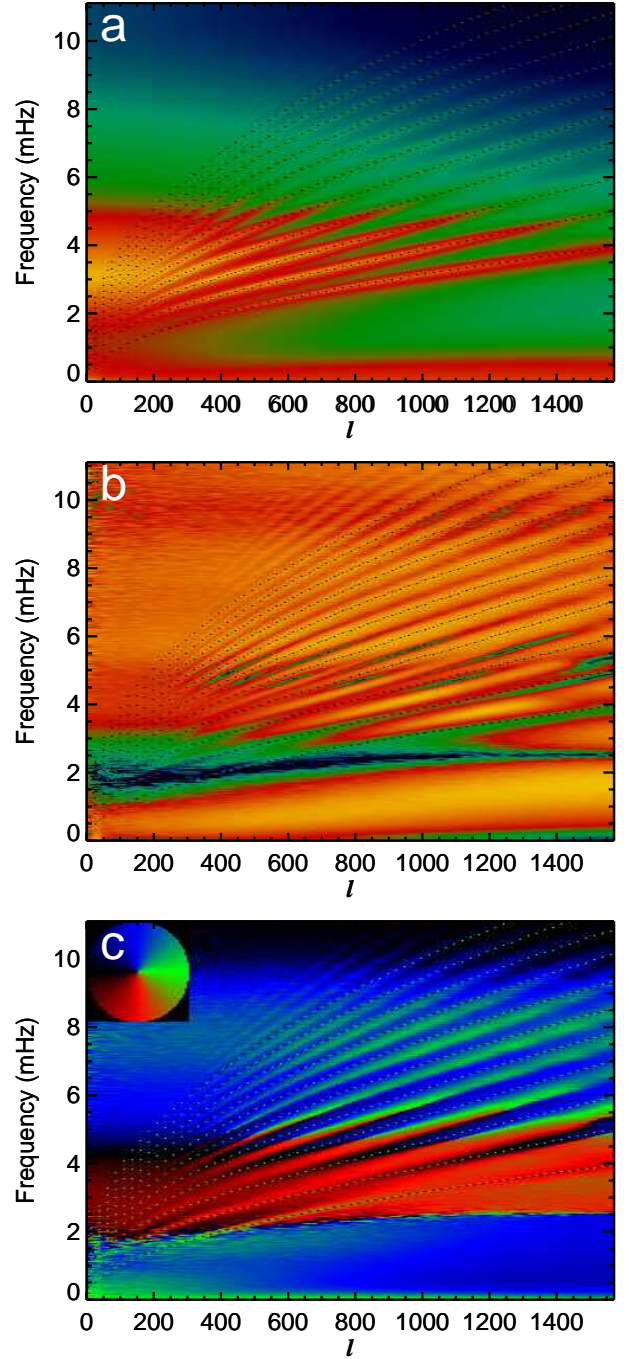}
\end{center}\caption[]{
From top to bottom, (a) shows the velocity
power on a logarithmic scale with a range of $10^6$ (blue-green-red-yellow),
(b) the magnitude of $\alpha$ on a logarithmic scale from 0.1 to 10 and
(c) the phase of $\alpha$.
The color coding for the phase is shown in the inset. Green ($0^\circ$)
corresponds to
the amplitude increasing with $I$, black ($180^\circ$) to decreasing,
blue ($90^\circ$) to the waves at higher intensity leading those at the reference intensity and
red ($270^\circ$) to trailing.
Rough mode frequencies (courtesy of R. Bogart) for radial order
$0 \le n \le 10$ are shown by dots.
}\label{pow1}\end{figure}

\section{Results}
\begin{figure}[t!]
\begin{center}
\includegraphics[width=0.9\columnwidth]{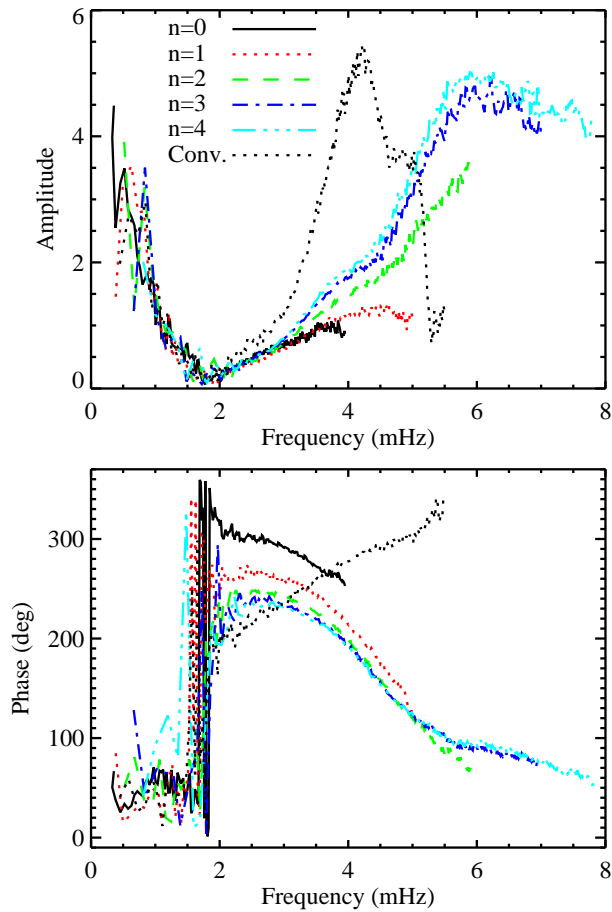}
\end{center}\caption[]{
Cuts through panels b and c of Fig. \ref{pow1} along the ridges for $n$=0 through 4. Dotted black line shows the results for the noise evaluated half
way between the $n=1$ and $n=2$ ridges.
For the phase convention see the caption to Fig. \ref{pow1}.
Note that the large oscillations in the phase just below 2 mHz are caused
by the very low amplitude in that region and thus large noise sensitivity.
This, in turn, also results in problems trying to unwrap the phase, which might
otherwise have improved the presentation.
}\label{cut2}\end{figure}

Fig. \ref{pow1} shows that there is a strong signal in both the amplitude
and phase of $\alpha$
and that the modes behave quite differently from the background
granulation signal.
To get a better picture, Figure \ref{cut2}
shows the amplitude and phase as a function of frequency along the ridges.
It is clear that both change dramatically as a function
of frequency. This is, perhaps, not surprising since the frequency determines
the near surface phase behavior of the modes \citep[see][]{2012ApJ...760L...1B}.

\begin{figure}[t!]
\begin{center}
\includegraphics[width=0.9\columnwidth]{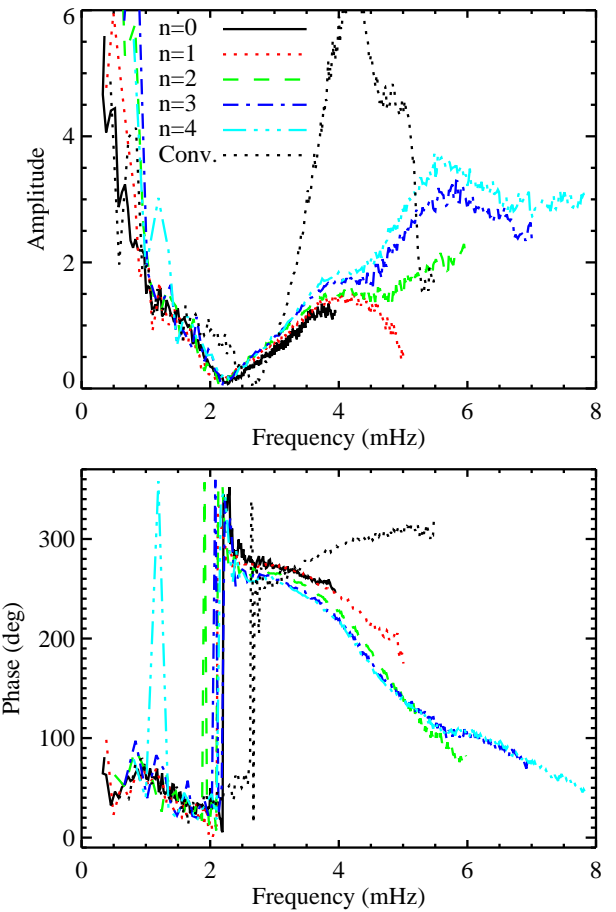}
\end{center}\caption[]{
As Fig. \ref{cut2}, but using intensities passed through a low pass filter
with a cutoff frequency of 2 mHz.
}\label{cut2f}\end{figure}

However, there is still significant scatter between the different radial orders,
especially for the phase.
Some of this is undoubtedly due to the varying oscillation/convection power
ratio, which in itself
makes the very low and high frequency results unreliable.
Another thing to consider is that the oscillations are visible in both $V$ and $I$.
Indeed, studying $V$ and $I$ together provides information on the mode
excitation and mode physics, as described in, e.g., \cite{2008SoPh..251..549S}
and \cite{2013SoPh..284..297S}.
This also means that some of the effects seen may be
due to the
visibility of the
modes in $I$, which is used to determine $S$.
To correct for this, $I^\prime$ is low pass filtered with a
cutoff of 2 mHz, to eliminate most of the mode signal and the
analysis repeated.
As Fig. \ref{cut2f} shows the
scatter of the phase is indeed significantly reduced.

The amplitudes are quite interesting. The change with intensity
is close to zero at the lowest reliable frequencies (i.e. $\approx 2.5$ mHz)
but becomes
larger than unity above about 3.5 mHz, in other words a unit change in background
intensity (corresponding, roughly to 100 km height change)
causes of order unity change in the amplitude in the observed velocity.

It is also worth nothing that the phases are generally not close
to 0 or 180 degrees. In other words the phases at different intensities are
not generally in phase.

These results are, perhaps, not surprising given the results of
\cite{2012ApJ...760L...1B}, who predicted large phase variations
with height. A quantitative comparison is quite difficult,
as \cite{2012ApJ...760L...1B} did not address the same problem
and since their final result depends on a delicate competition between
positive and negative contributions from different heights
and details of the radiative transfer.
But noting that a unit change in the relative intensity corresponds to
roughly 100 km height difference and assuming that there are no other effects,
they saw phase changes of order tens of seconds (of order a radian),
so it is not surprising that we observe changes of order unity.
However, more detailed predictions would require more detailed
theoretical work. For attempts at this the reader is referred to
\cite{2004A&A...418..305S} and references therein.
Similarly, substantial insight might be obtained from numerical
simulations.

As for the robustness of the results, it may be noted that using the low pass
filtered Doppler velocity instead of the intensity leads to very similar
results, except for a sign change (velocity and intensity are strongly
anti-correlated in the granulation).
Also, assuming that the variation is linear with intensity and that the
velocity measurement process is linear, it follows that
the slope is insensitive to blurring and thus to
the instrument PSF.

\begin{figure}[t!]
\begin{center}
\includegraphics[width=0.9\columnwidth]{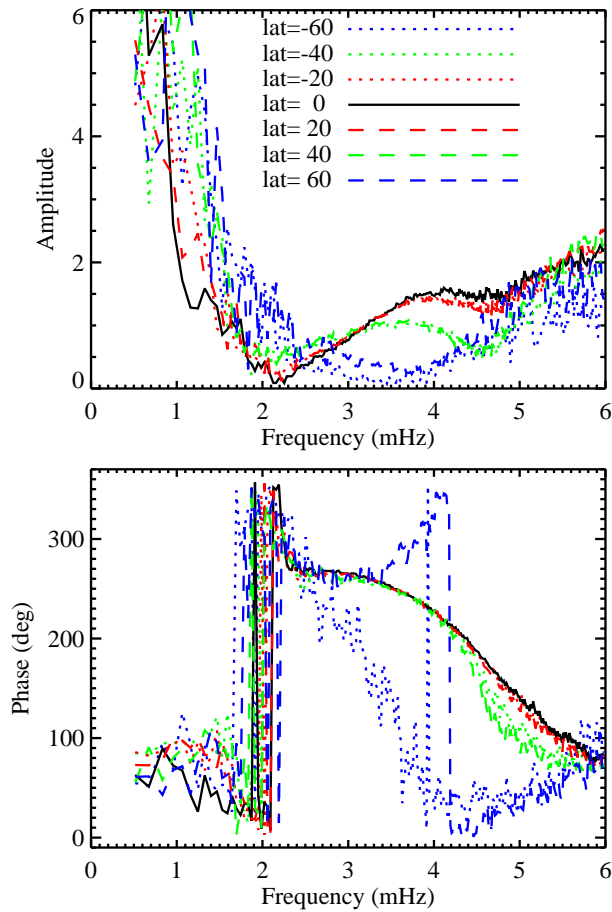}
\end{center}\caption[]{
Amplitudes and phase shifts for $n=2$ at various latitudes. Note that the average
B0 angle was only about $1^\circ$, so the viewing angle is effectively identical
to the latitude.
}\label{cut_lat}\end{figure}

As a likely cause of the phase shifts is the height dependence of the phase of
the oscillations combined with the varying height of observation, one might
expect some change with viewing angle, as that also changes the effective
height of observation. As seen from Fig. \ref{cut_lat} the phases
are almost unaffected up to $40^\circ$ from disk center.
At $60^\circ$ the results are dramatically changed, but are different
for $+60^\circ$ and $-60^\circ$, so this is likely not a robust result,
most likely due to the substantial foreshortening.
The amplitudes show more consistent changes. To what extent
those changes are due to changes in the S/N has
not been investigated.

\begin{figure}[t!]
\begin{center}
\includegraphics[width=0.9\columnwidth]{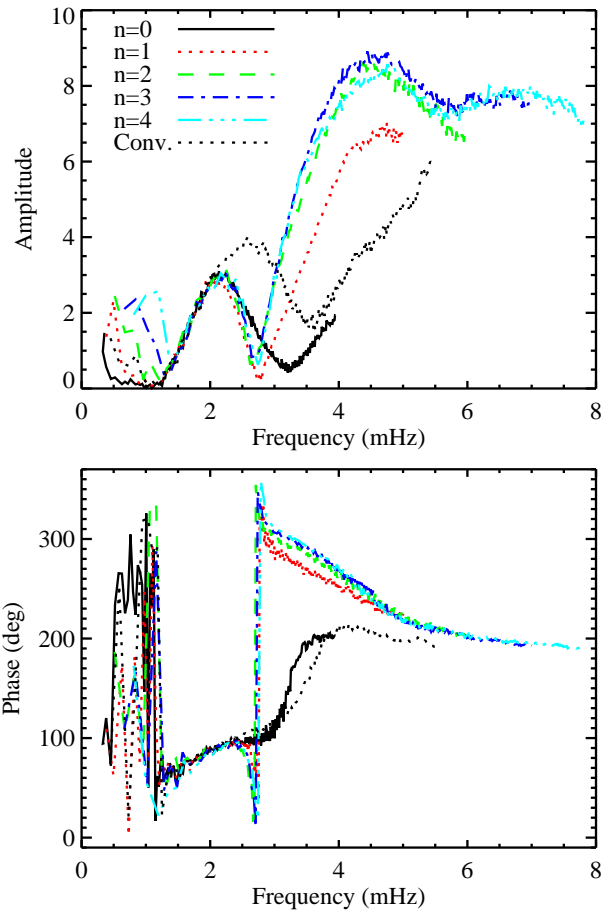}
\end{center}\caption[]{
As Fig. \ref{cut2f} but for the intensity.
}\label{cut2fi}\end{figure}

One might also expect the intensity signal of the oscillation to change with
position within the granulation pattern, which as Fig. \ref{cut2fi}
shows is indeed the case. Here the raw intensities were regressed with the filtered
ones, in the same way as was done for the velocities.
Given the lower S/N for intensity, these
results are probably more affected by noise, but it is still clear that
there is a strong signal and while the results for the modes and the
background appear similar around 3 mHz, the S/N is actually quite significant
there.
It is interesting that the
results are substantially more $n$ dependent, probably reflecting
differences in the mode physics (e.g. f modes do not compress
the matter significantly, while p modes do).
It is also interesting that the frequency dependence is different from that
of the velocity, presumably reflecting the different heights of
formation.
It has not been investigated how other variables, such as line depth or width, vary.

Finally it may be noted that previous studies of waves over granules and
intergranular lanes have been made (see, for example, \cite{2001A&A...369..660K} and \cite{2006ARep...50..588K} and references therein). Those studies were
mostly concerned with attempts at detecting the excitation of
modes or interactions with magnetic fields. Also, they did not
investigate the dependence on $k$ and $\omega$ and so were not able to resolve
the modes clearly, which makes comparisons with the present work difficult.

\section{Conclusion}
Using the method described it is shown that
the amplitudes and phases of the oscillations
depend strongly on position in the granulation.
Of course, this represents a correlation and does not imply
that the intensity is the proximate cause. The velocity and intensity
are highly correlated and no attempt was made to distinguish between
their effects.
No physical model of the effect has been made, but the
large phase change with geometric height reported by \cite{2012ApJ...760L...1B}
combined with the change of observing height with intensity in the granulation
is a promising candidate.
Also, instrumental effects can not be completely excluded,
especially for the amplitude effect, as there is some change
in Doppler sensitivity with
the thermodynamic properties where the line is formed.

A promising way to investigate the origin of the effects reported here is
to analyze the results of large scale numerical simulations of near
surface convection.
Indeed, studies similar to those of \cite{2013SoPh..284..297S}, but for
the effects studied here, may prove very useful.

Observations with higher spatial resolution may also help shed more
light on the relevant processes. In particular it may be possible to
look for non-linearities and to disentangle the intensity and velocity
effects. Similarly, higher spectral resolution may allow for the height
dependence to be determined.
It may also be possible to correlate against other variables, such as
the magnetic field strength, alone or in combination with the intensity,
thereby gaining further insight about e.g. magnetic field effects,
a subject of significant current interest.

Note that,
similar to the analysis by \cite{2014SoPh..289.3457N}, the analysis here only
relies on existing HMI observations and does
not require new observations using higher spatial resolution or
other spectral lines. It is thus possible to study and exploit the
effects described here using the existing 5 years of HMI data.
An interesting possibility is that the effects described here
may be exploited 
to further separate granulation from oscillations, thereby possibly 
leading to an improved S/N ratio.

\begin{acknowledgements}
The author would like to thank Aaron Birch, Tom Duvall and Björn Löptien
for useful discussions.
The HMI data are courtesy of NASA/{\it SDO} and the HMI science team.
The data were processed at the German Data Center for SDO (GDC-SDO),
funded by the German Aerospace Center
(DLR). Support is acknowledged from the SPACEINN and SOLARNET projects of
the European Union.
\end{acknowledgements}


\bibliographystyle{aa}
\bibliography{vint}

\begin{thebibliography}{16}
\expandafter\ifx\csname natexlab\endcsname\relax\def\natexlab#1{#1}\fi

\bibitem[{{Baldner} \& {Schou}(2012)}]{2012ApJ...760L...1B}
{Baldner}, C.~S. \& {Schou}, J. 2012, \apjl, 760, L1

\bibitem[{{Ball} \& {Gizon}(2014)}]{2014A&A...568A.123B}
{Ball}, W.~H. \& {Gizon}, L. 2014, \aap, 568, A123

\bibitem[{{Bhattacharya} {et~al.}(2015){Bhattacharya}, {Hanasoge}, \&
  {Antia}}]{2015ApJ...806..246B}
{Bhattacharya}, J., {Hanasoge}, S., \& {Antia}, H.~M. 2015, \apj, 806, 246

\bibitem[{{Christensen-Dalsgaard} {et~al.}(1988){Christensen-Dalsgaard},
  {Dappen}, \& {Lebreton}}]{1988Natur.336..634C}
{Christensen-Dalsgaard}, J., {Dappen}, W., \& {Lebreton}, Y. 1988, \nat, 336,
  634

\bibitem[{{Hanasoge} {et~al.}(2013){Hanasoge}, {Gizon}, \&
  {Bal}}]{2013ApJ...773..101H}
{Hanasoge}, S.~M., {Gizon}, L., \& {Bal}, G. 2013, \apj, 773, 101

\bibitem[{{Khomenko} {et~al.}(2001){Khomenko}, {Kostik}, \&
  {Shchukina}}]{2001A&A...369..660K}
{Khomenko}, E.~V., {Kostik}, R.~I., \& {Shchukina}, N.~G. 2001, \aap, 369, 660

\bibitem[{{Kostyk} {et~al.}(2006){Kostyk}, {Shchukina}, \&
  {Khomenko}}]{2006ARep...50..588K}
{Kostyk}, R.~I., {Shchukina}, N.~G., \& {Khomenko}, E.~V. 2006, Astronomy
  Reports, 50, 588

\bibitem[{{Langfellner} {et~al.}(2014){Langfellner}, {Gizon}, \&
  {Birch}}]{2014A&A...570A..90L}
{Langfellner}, J., {Gizon}, L., \& {Birch}, A.~C. 2014, \aap, 570, A90

\bibitem[{{Nagashima} {et~al.}(2014){Nagashima}, {L{\"o}ptien}, {Gizon},
  {Birch}, {Cameron}, {Couvidat}, {Danilovic}, {Fleck}, \&
  {Stein}}]{2014SoPh..289.3457N}
{Nagashima}, K., {L{\"o}ptien}, B., {Gizon}, L., {et~al.} 2014, \solphys, 289,
  3457

\bibitem[{{Piau} {et~al.}(2014){Piau}, {Collet}, {Stein}, {Trampedach},
  {Morel}, \& {Turck-Chi{\`e}ze}}]{2014MNRAS.437..164P}
{Piau}, L., {Collet}, R., {Stein}, R.~F., {et~al.} 2014, \mnras, 437, 164

\bibitem[{{Schou} {et~al.}(2012){Schou}, {Scherrer}, {Bush}, {Wachter},
  {Couvidat}, {Rabello-Soares}, {Bogart}, {Hoeksema}, {Liu}, {Duvall}, {Akin},
  {Allard}, {Miles}, {Rairden}, {Shine}, {Tarbell}, {Title}, {Wolfson},
  {Elmore}, {Norton}, \& {Tomczyk}}]{2012SoPh..275..229S}
{Schou}, J., {Scherrer}, P.~H., {Bush}, R.~I., {et~al.} 2012, \solphys, 275,
  229

\bibitem[{{Severino} {et~al.}(2013){Severino}, {Straus}, {Oliviero}, {Steffen},
  \& {Fleck}}]{2013SoPh..284..297S}
{Severino}, G., {Straus}, T., {Oliviero}, M., {Steffen}, M., \& {Fleck}, B.
  2013, \solphys, 284, 297

\bibitem[{{Severino} {et~al.}(2008){Severino}, {Straus}, \&
  {Steffen}}]{2008SoPh..251..549S}
{Severino}, G., {Straus}, T., \& {Steffen}, M. 2008, \solphys, 251, 549

\bibitem[{{Snodgrass}(1984)}]{1984SoPh...94...13S}
{Snodgrass}, H.~B. 1984, \solphys, 94, 13

\bibitem[{{Stix} \& {Zhugzhda}(2004)}]{2004A&A...418..305S}
{Stix}, M. \& {Zhugzhda}, Y.~D. 2004, \aap, 418, 305

\bibitem[{{Zhao} {et~al.}(2012){Zhao}, {Nagashima}, {Bogart}, {Kosovichev}, \&
  {Duvall}}]{2012ApJ...749L...5Z}
{Zhao}, J., {Nagashima}, K., {Bogart}, R.~S., {Kosovichev}, A.~G., \& {Duvall},
  Jr., T.~L. 2012, \apjl, 749, L5

\end{thebibliography}

\end{document}